\documentclass[prd,aps,twocolumn,showpacs,nofootinbib]{revtex4-1}

\usepackage{color}

\definecolor{green}{rgb}{0,.5,0}

\definecolor{red}{rgb}{1,0,0}

\usepackage{graphicx}
\usepackage[subfigure]{graphfig}
\usepackage{epsfig}
\usepackage[colorlinks,linkcolor=red,anchorcolor=green,citecolor=blue,CJKbookmarks=True]{hyperref}
\usepackage{dcolumn}
\usepackage{amsmath,amssymb,amsthm}
\usepackage{latexsym}
\usepackage{multirow}
\usepackage{bbm}
\usepackage{ulem}
\def\be{\begin{equation}}
\def\ee{\end{equation}}
\def\bea{\begin{eqnarray}}
\def\eea{\end{eqnarray}}

\def\ln{\mathrm{ln}}
\def\diag{\mathrm{diag}}

\def\tr{\mathrm{tr}}
\begin{document}

\title{Glueballs at Physical Pion Mass}

\author{Feiyu Chen$^{1,2}$, Xiangyu Jiang$^{1,2}$, Ying Chen$^{1,2}$, Keh-Fei Liu$^3$, Wei Sun$^{1}$, Yi-Bo Yang$^{2,4,5,6}$
\vspace*{-0.5cm}
\begin{center}
\large{
\vspace*{0.4cm}
\includegraphics[scale=0.15]{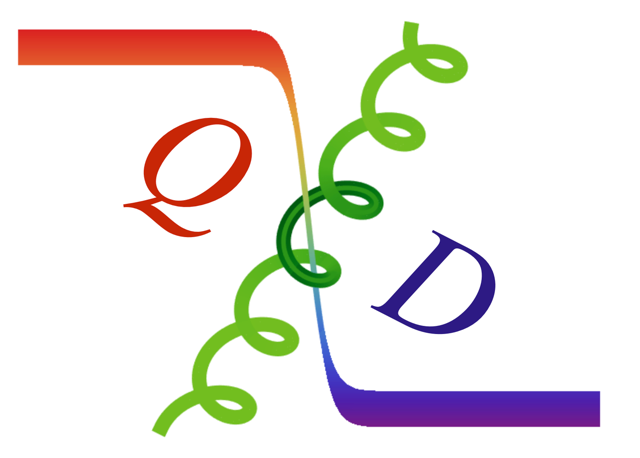}\\
\vspace*{0.4cm}
($\chi$QCD Collaboration)
}
\end{center}
}
\affiliation{
$^{1}$\mbox{Institute of High Energy Physics, Chinese Academy of Sciences, Beijing 100049, China}\\
$^{2}$\mbox{School of Physics Sciences, University of Chinese Academy of Sciences, Beijing 100049, China}\\
$^{3}$\mbox{Department of Physics and Astronomy, University of Kentucky, Lexington, Kentucky 40506, USA}\\
$^{4}$\mbox{CAS Key Laboratory of Theoretical Physics, Institute of Theoretical Physics, Chinese Academy of Sciences, Beijing 100190, China}\\
$^{5}$\mbox{School of Fundamental Physics and Mathematical Sciences, Hangzhou Institute for Advanced Study, UCAS, Hangzhou 310024, China}\\
$^{6}$\mbox{International Centre for Theoretical Physics Asia-Pacific, Beijing/Hangzhou, China}\\
}

\begin{abstract}
We study glueballs on two $N_f=2+1$ RBC/UKQCD gauge ensembles with physical quark masses at two lattice spacings. The statistical uncertainties of the glueball correlation functions are considerably reduced through the cluster decomposition error reduction (CDER) method. The Bethe-Salpeter wave functions are obtained for the scalar, tensor and pseudoscalar glueballs by using spatially extended glueball operators defined through the gauge potential $A_\mu(x)$ in the Coulomb gauge. These wave functions show similar features of non-relativistic two-gluon systems, and they are used to optimize the signals of the related correlation functions at the early time regions.
Consequently, the ground state masses can be extracted precisely. To the extent that the excited state contamination is not important, our calculation gives glueball masses at the physical pion mass for the first time. 
\end{abstract}

\maketitle

\textit{Introduction} 
In quantum chromodynamics (QCD), gluons carry color charges and have the direct strong interaction among themselves. Therefore it is expected that hadrons can be made up of gluons solely, namely glueballs. The property of glueballs has been one of the hottest topics in hadron physics for more than several decades and have aroused extensive and intensive experimental and theoretical studies.  However, their existence has not been confirmed yet. 

Glueballs are well defined objects in the quenched approximation where the quark-gluon transitions are switched off, and the quenched lattice QCD (QLQCD) studies have confirmed the existence of purge gauge glueballs and have predicted their spectrum~\cite{Morningstar:1999rf,Chen:2005mg}. In the presence of dynamical quarks, the situation is more complicated owing to the decays of glueballs and the possible mixing between glueballs and $q\bar{q}$ mesons. There have been a few preliminary full-QCD lattice studies~\cite{Richards:2010ck,Gregory:2012hu,Sun:2017ipk} at pion masses much larger than the physical value, which have observed possible glueball states with masses close to the predictions from QLQCD. However, it is still
challenging to verify the previous glueball studies at the physical pion mass due to the computational cost, since glueball relevant lattice studies usually require large statistics of thousands gauge configurations, which is computationally prohibitive for physical pion masses and large spatial volumes in the present stage. Fortunately, the cluster decomposition principle ensures that the correlation length between the glueball operators is insensitive to the volume, and can be implemented to reduce the statistical errors of glueball correlation functions. In Ref.~\cite{Sun:2015enu,Liu:2017man,Yang:2018bft}, a cluster decomposition error reduction (CDER) method was introduced to suppress the statistics requirement on lattices with large spatial volumes. In this work, we use the CDER method to trade the statistical uncertainty for negligible systematic one, and obtain clear mass spectrum and Bethe-Salpeter wave function with the pure glueball operators on only $\sim$ 300 configurations at physical pion mass.


\begin{table}[t]
	\centering \caption{\label{tab:lattice}	 Parameters of 48I and 64I ensemble~\cite{RBC:2014ntl}.}
	\begin{ruledtabular}
		\begin{tabular}{ccccc}
			$L^3\times T$   &  $a^{-1}$ (GeV)  &  $m_\pi$ (MeV) & $La$ (fm) & $N_\mathrm{conf}$ \\
			$48^3\times\ \ \! 96$ & 1.730(4) & $139.2(4)$ & $5.476(12)$    & 364\\
			$64^3 \times 128$& 2.359(7) & $139.2(5)$ & $5.354(16)$  &  300\\
		\end{tabular}
	\end{ruledtabular}
\end{table}
\textit{Numerical details} 
We choose two RBC/UKQCD  gauge ensembles (labeled as 48I and 64I) with $2+1$ flavor domain wall fermion at physical pion and kaon masses and with spatial sizes around $5.5$ fm. The parameters of the ensembles are shown in Table~\ref{tab:lattice} of reference~\cite{RBC:2014ntl}. In order to extract glueball states, we adopt the same strategy in Ref.~\cite{Morningstar:1999rf,Chen:2005mg} to build the glueball operator set $\mathcal{S}(R^{PC})=\{\mathcal{O}_\alpha, \alpha=1,2,\ldots, 24\}$ for the scalar ($R^{PC}=A_1^{++}$), pseudoscalar ($A_1^{-+}$), and tensor ($E^{++}\oplus T_2^{++}$) channels, where $R=A_1, E, T_2$ are the irreducible representations of the lattice symmetry group $O$, $P$ and $C$ are the parity and the charge conjugate quantum numbers, respectively. 

It is well known that the glueball relevant lattice study usually requires a large statistics, but we have only $N_\mathrm{conf}\sim 300$ configurations available for the two ensembles. Fortunately, its lattice size is large enough to make the cluster decomposition error reduction method (CDER) proposed in Ref.~\cite{Liu:2017man} to be efficient on improving the signal-to-noise ratio in the calculation. The idea of CDER is the following: since $\mathcal{C}_{\alpha\beta}(\mathbf{r},t)\equiv \langle 0|\mathcal{O}_\alpha(\mathbf{r},t)\mathcal{O}_\beta(\mathbf{0},0)|0\rangle$ behaves as $\sim \xi^{-\frac32} e^{-m\xi}$ for large $\xi$, where $m$ is the lowest mass in this channel and $\xi=(\mathbf{r}^2+t^2)^{1/2}$ is the Euclidean separation between the sink and the source operator, the signals of $\mathcal{C}_{\alpha\beta}$ will be undermined by statistical noises when $|\mathbf{r}|$ is beyond some length scale $|\mathbf{r}_c|\propto 1/m$. Therefore the correlation matrix $C_{\alpha\beta}$ with the operator set $\mathcal{S}(R^{PC})$ in a given $R^{PC}$ channel can be calculated as 
\begin{eqnarray}\label{eq:estimate}
\mathcal{C}_{\alpha\beta}(t)&=&\frac{1}{T}\sum\limits_{\tau,\mathbf{x,r}} \langle 0|\mathcal{O}_\alpha(\mathbf{x+r},t+\tau)\mathcal{O}_\beta(\mathbf{x},\tau)|0\rangle\nonumber\\
&\simeq &\sum\limits_{|\mathbf{r}|<r_c} K_{\alpha\beta}(\mathbf{r},t)\equiv \mathcal{C}_{\alpha\beta}(r_c,t),
\end{eqnarray} 
where the average over time slices is taken into account to improve the statistics and the kernel functions $K_{\alpha\beta}(\mathbf{r},t)$ is introduced as  
\begin{equation}\label{eq:kernel}
    K_{\alpha\beta}(\mathbf{r},t)=\sum\limits_{\mathbf{k}} e^{-i\mathbf{k}\cdot \mathbf{r}} \hat{\mathcal{O}}_\alpha(-\mathbf{k})\hat{\mathcal{O}}_\beta(\mathbf{k},t)
\end{equation}
in terms of the Fourier transformed operators $\hat{\mathcal{O}}_\alpha(-\mathbf{k},t)=\sum\limits_{\mathbf{x}}e^{-i\mathbf{k}\cdot\mathbf{x}}\mathcal{O}_\alpha(\mathbf{x},t)$.  The cutting scale parameter $r_c$ is chosen empirically when the value of $\mathcal{C}_{\alpha\beta}(r_c,t)$ saturates. 
 
\begin{figure}[t]
	\centering
	\includegraphics[height=6.0cm]{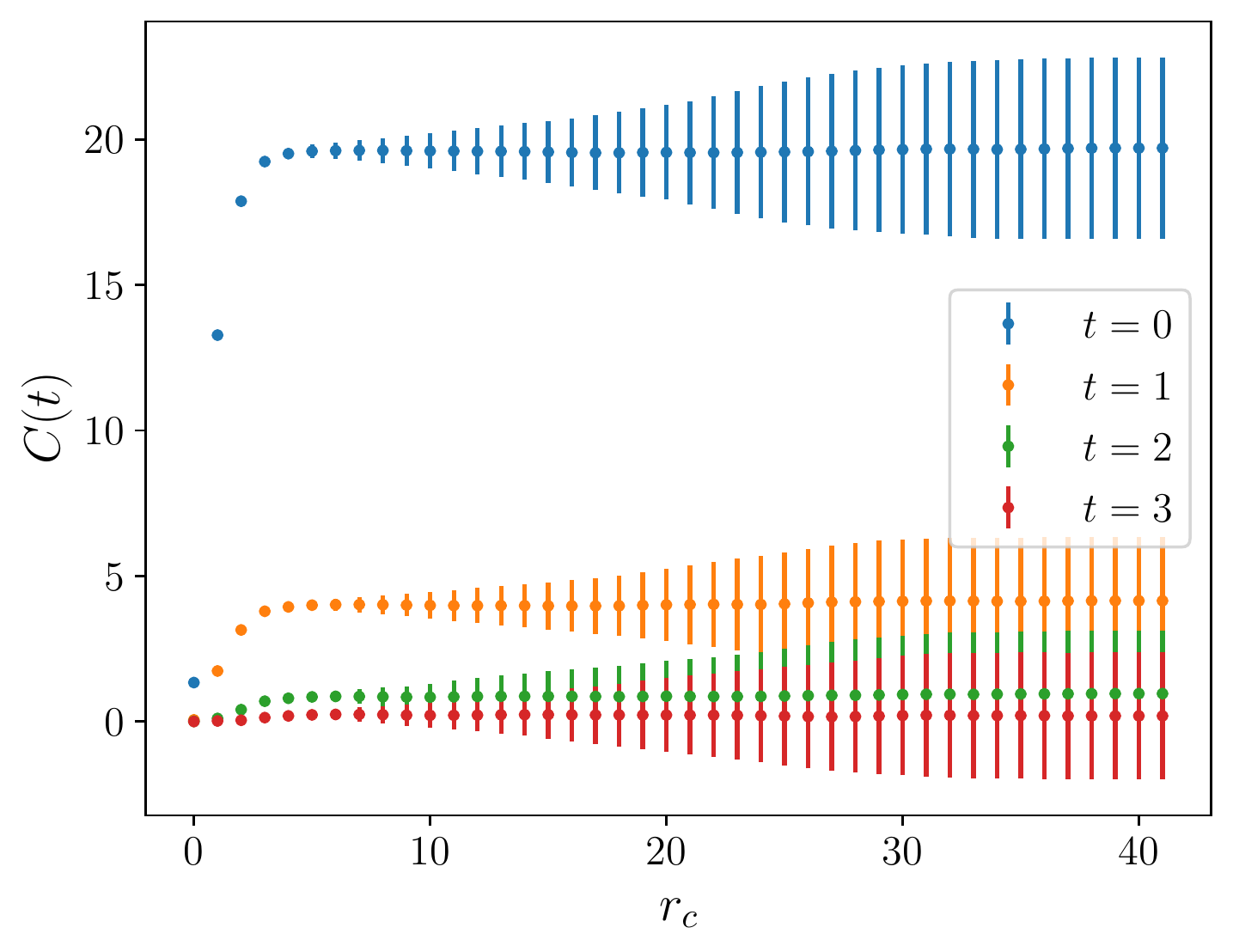}
	\caption{Plot of $A_1^{++}$ correlation function at first 4 times lices against the cutoff on $48^3 \times 64$ lattice. 
		The $r_c$ is in lattice unit. As expected, the signal reaches maximum at $r\sim 7a$ and only noise is increasing
		afterwards. Here the errorbars indicate the standard deviations instead of 
		statistic errors.}
	\label{fig:a1pp_cut_plot}
\end{figure}

 The efficacy of CDER is illustrated in Fig.~\ref{fig:a1pp_cut_plot}, where a typical $\mathcal{C}_{\alpha\beta}(r_c,t)$ in $A_1^{++}$ channel at $t/a=0,1,2,3$ is plotted versus $r_c/a$, where we use the standard deviations instead of the statistical ones such that the varying of errors with respect to $r_c$ to be seen more clearly. One can see that the central values of $\mathcal{C}_{\alpha\beta}(r_c,t)$ at different $t$ saturate uniformly beyond $r_c/a=7$ but the errors grow when increasing $r_c$. In the practical calculation we choose $r_c/a=7$ for all the $\mathcal{C}_{\alpha\beta}(R,t)$ in all the channels. 

In each $R^{PC}$ channel, after the correlation matrix $\mathcal{C}_{\alpha\beta}(t)$ is estimated through Eq.~(\ref{eq:estimate}), we perform the standard variational method by solving the generalized eigenvalue problem $\mathcal{C}_{\alpha\beta}(t_0)v_\beta=\lambda  \mathcal{C}_{\alpha\beta}(0)v_\beta$, where we choose $t_0/a=1$. The eigenvector $v_\alpha^{(n)}$ of the $n$-th largest eigenvalue gives an optimized operator $\mathcal{O}_n=v_\alpha^{(n)}\mathcal{O}_\alpha$ that is expected to couple most with the $n$-th lowest state $|n\rangle$ contributing to $\mathcal{C}_{\alpha\beta}(t)$. In practice, the eigenvectors $v_\alpha^{(n)}$ are normalized by $\mathcal{C}_n(t=0)=\langle 0| \mathcal{O}_n(0)\mathcal{O}_n(0)|0\rangle=1$, which implies $\langle 0|\mathcal{O}_n|m\rangle \approx \delta_{mn}$ if the states are normalized as $\langle m|n\rangle=\delta_{mn}$. 


Although the operator basis in each channel is large, it is found that the correlation function of the operator $\mathcal{O}_1$ optimized for the ground state has still a sizeable contribution from higher states, (the effective mass function does not show
a plateau good enough in the early time range, see Fig.~\ref{fig:GG_mass} in the Supplemental materials~\cite{sm}).  In order to enhance the contribution of the ground state to the corresponding correlation functions, we construct another type of gluonic operators in terms of the gauge potential $A_\mu(x)$~\cite{deForcrand:1991kc,Liang:2014jta} in the Coulomb gauge. 
According to the definition of the gauge link oriented in the $\mu$ direction starting from the lattice site, i.e, $U_\mu(x)= \exp(-igA_\mu(x+a\hat{\mu}/2))$, up to an irrelevant pre-factor we have 
\begin{eqnarray}
A_\mu(x) \equiv \ln U_\mu(x+a\hat{\mu})-\ln U_\mu(x-a\hat{\mu}) 
\end{eqnarray}
where $\ln U_\mu(x)$ can be derived by $\ln U_\mu(x)= V(x) \diag(\ln \lambda_1(x), \ln \lambda_2(x), \ln\lambda_3(x)) V^\dagger(x)$ with $\lambda_{1,2,3}(x)$ being the eigenvalues of $U_\mu(x)$ and $V(x)$ being the unitary matrix that diagonalizes $U_\mu(x)$. On each time slice, the spatially extended operator $O_A^R(r,t)$ can be constructed by $A_\mu(x)$ at two different points. For $P=C=+$, the explicit expression of $O_A^R(r,t)$ of the quantum number $R^{PC}$ is 
\begin{equation}
O_A^R(\mathbf{x},t;r)=\frac{1}{N_r} \sum\limits_{ij,|\mathbf{r}|=r} \tr S_{ij}^R[A_i(\mathbf{x+r})A_j(\mathbf{x})]
\end{equation} 
where $N_r$ is the degeneracy of $\mathbf{r}$ with the same $r$ and $S_{ij}^R$ are the combinational coefficients related to the irreducible representation $R$ of the spatial symmetry group $O$ (octahedral group) on the lattice. For $R=A_1$, the non-zero values of $S_{ij}^R$ are $S_{11}^R=S_{22}^R=S_{33}^R=1$. The $E$ representation has two components, the first of which is given by $S_{11}^R=-S^R_{22}=1/\sqrt{2}$, and the other is given by $S_{11}^R=S_{22}^R=-1/\sqrt{6}$ and $S_{33}^R=2/\sqrt{6}$. The three components of $T_2$ have $S_{ij}^{R,k}=|\epsilon_{ijk}|$. The $O_A^R(r,t)$ operator for $A_1^{-+}$ is defined as 
\begin{equation}
O_A^{A_1^{-+}}(\mathbf{x},t;r)=\frac{1}{N_r}\sum_{|\vec r| = r} \epsilon_{ijk} \tr\left[A_i(\vec x + \vec r) A_j(\vec x)\right] \hat r_k
\end{equation}
where $\hat{\mathbf{r}}$ is the orientation vector of $\mathbf{r}$.

\begin{figure}[t!]
	\centering
    \includegraphics[width=0.46\textwidth]{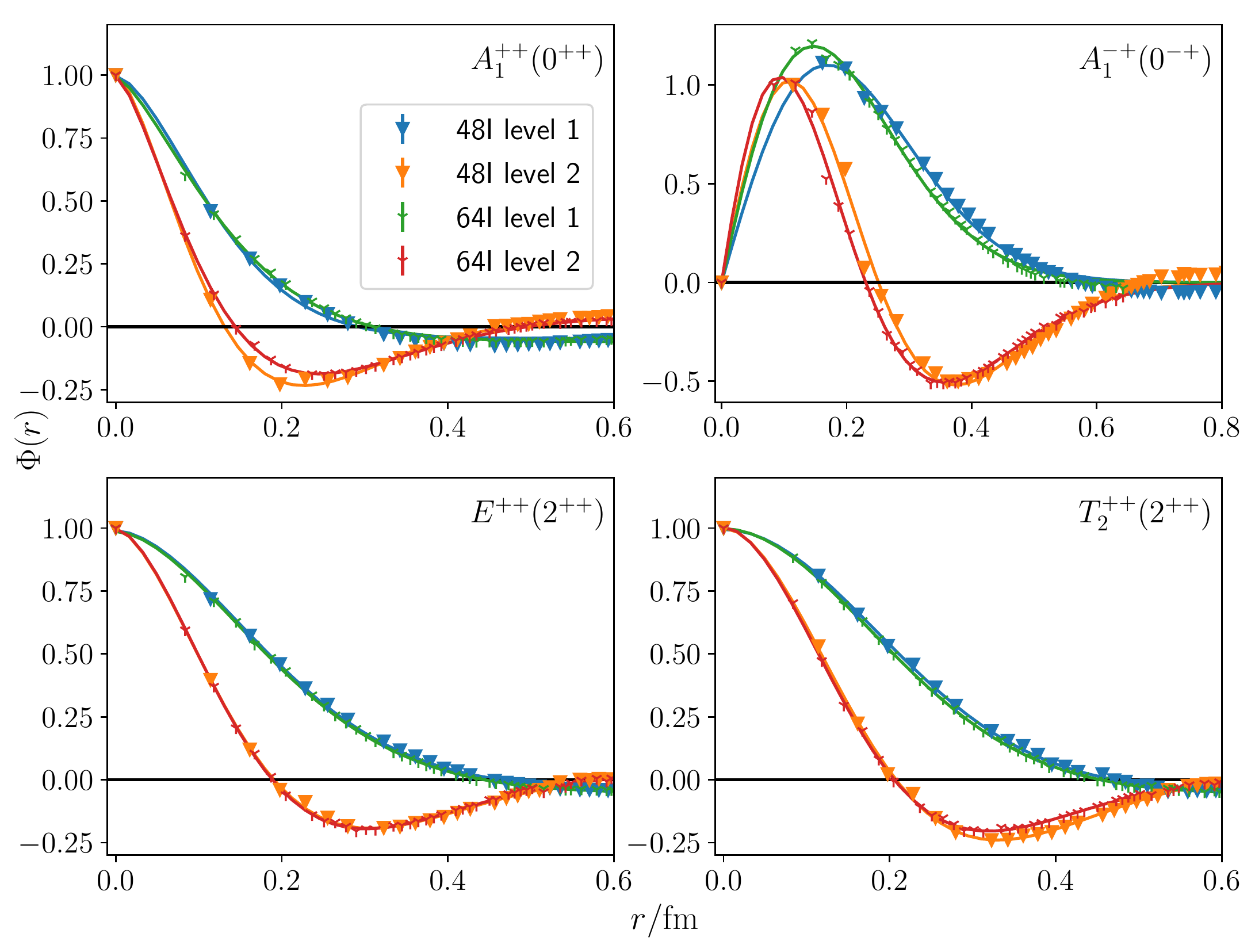}
	\caption{Normalized BS wave functions of ground and first excited states on 48I and 64I lattice.}
	\label{fig:bs-wavefunc}
\end{figure}

In each channel, by the implementation of CDER similar to Eq.~(\ref{eq:estimate}) and ~(\ref{eq:kernel}), we calculate the correlation function $\mathcal{C}_{An}(r,t)$ of $O_A^R$ and the $n$-th optimized operator $\mathcal{O}_n$,
\begin{eqnarray}\label{eq:aa-corr}
\mathcal{C}_{An}(r,t)&=&\frac{1}{T}\sum\limits_{\tau,\mathbf{x,y}} \langle 0|\mathcal{O}_A^R(\mathbf{x},t+\tau;r)\mathcal{O}_n(\mathbf{y},\tau)|0\rangle\nonumber\\
&\approx& \Phi_n(r)e^{-m_n t}+\sum\limits_{i\ne n}\epsilon_m \Phi_m(r)e^{-m_i t},
\end{eqnarray} 
where the last parameterization is due to predominant coupling of $\mathcal{O}_n$ to the $n$-th state with $\epsilon_i \ll 1$ and $\Phi_n(r)\propto \langle 0|O_A^R(\mathbf{0},t;r)|n\rangle$ is interpreted as the Bethe-Salpeter (BS) wave function~\cite{bib:bethe_salpeter} of the $n$-th state $|n\rangle$ if it is a two-gluon glueball state.
In principle, $\Phi_n(r)$ can be approximated by $\mathcal{C}_{An}(r,t)$ at any $t$ according to Eq.~(\ref{eq:aa-corr}). However, due to the rapid increase of the noises, we can only observe clear signals of $\Phi_n(r)$ at the first few time slices. Figure~\ref{fig:bs-wavefunc} shows the profiles of $\Phi_{1,2}(r)$ at $t=0$ for $A_1^{++}$, $A_1^{-+}$, $E^{++}$ and $T_2^{++}$ channels on 48I and 64I lattices (normalized by $\Phi(r=0) = 1$ for $PC=++$ channels and $\Phi(r=a)$ for the $A_1^{-+}$ channel), where the values of $r$ are converted to physical units using the lattice spacings $a$ in Table~\ref{tab:lattice}. The data points are the calculated values and the curves are plotted using the fitted parameters through some phenomenological functional forms (see the Supplemental materials~\cite{sm}).  
It is interesting that the $r$-behaviors of $\Phi_{1,2}(r)$ in $PC=++$ channels are very similar to the radial $1S$ and $2S$ Schr\"{o}dinger wave function of a two-body system with a central potential, while those in the $A_1^{-+}$ channel have the feature of $1P$ and $2P$ wave functions. Note that by solving the Bethe-Salpeter equation of the $0^{-+}$ glueball, the two-body $P$-wave feature of the $0^{-+}$ Bethe-Salpeter amplitude was also observed in Ref.~\cite{Souza:2019ylx}. 

\begin{figure*}[t]
	\centering
	\includegraphics[height=12cm]{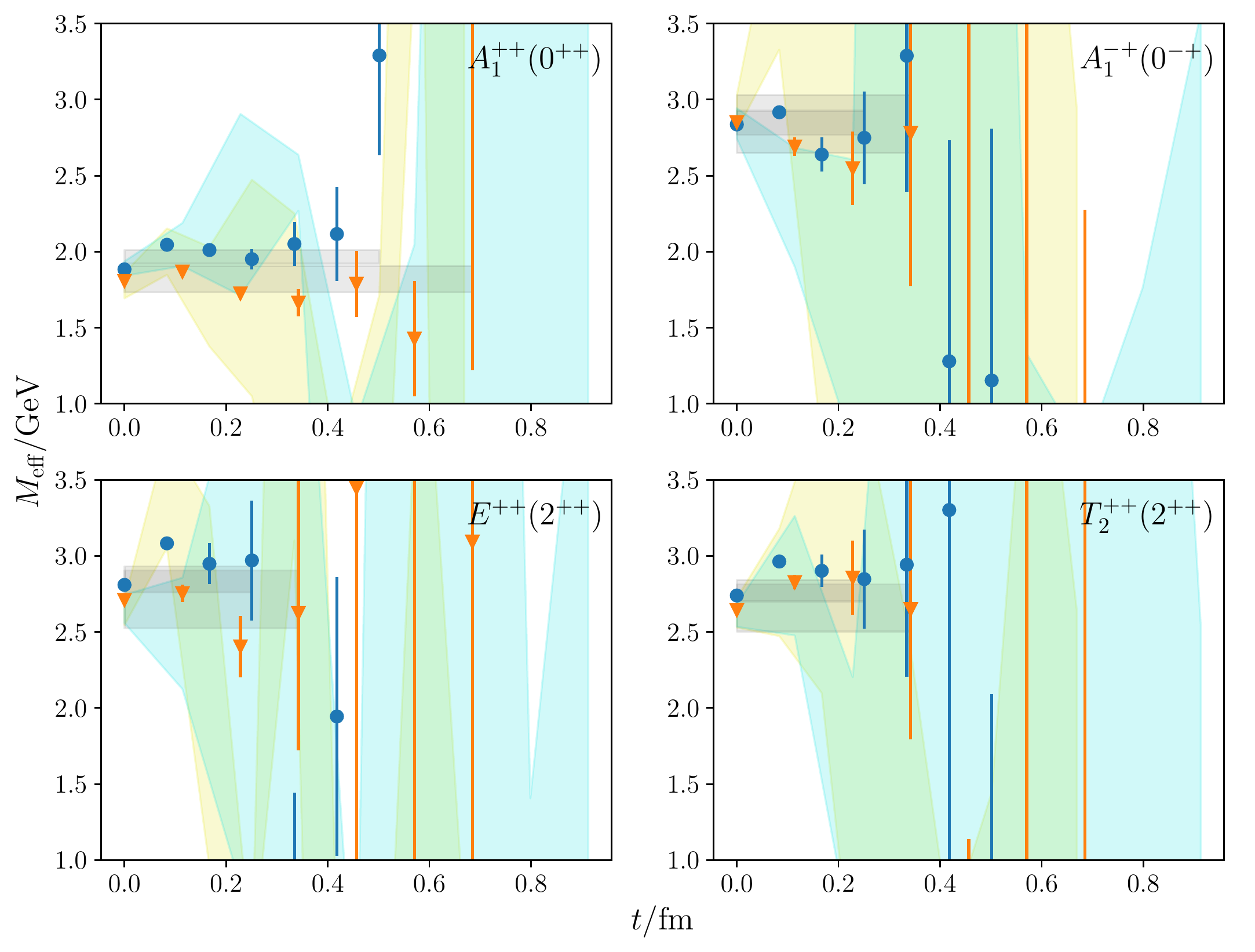}
	\caption{Effective masses of $C_{A1}(r_1,t)$ on 48I (orange triangles) and 64I (blue dots), with shaded region showing the results without CDER. The gray bands are the fit results through single-exponential function forms.}
	\label{fig:final_mass_plat}
\end{figure*}

 We would not like to put too much emphasis on the interpretation of the wave functions $\Phi_n(r)$, but will use the feature of the $r$-behaviors of these functions to enhance the contribution of the ground states in the early time region. Since $\Phi_2(r)$ in each channel has a radial node at $r=r_1$, the correlation function $\mathcal{C}_{A1}(r_1,t)$ in each channel is expected to be dominated by the ground state, since the second term in Eq.~(\ref{eq:aa-corr}) is further suppressed by $\Phi_2(r_1)\approx 0$. According to Fig.~\ref{fig:bs-wavefunc}, the radial node of $\Phi_2(r)$ appears at $r\sim 0.12, 0.19$ and 0.22 fm in the scalar ($A_1^{++}$), the tensor ($E^{++}$ and $T_2^{++}$) and the pseudoscalar channel ($A_1^{-+}$), respectively. According to the lattice spacings in Table~\ref{tab:lattice}, we choose $r_1(A_1^{++})/a=1 $, $r_1(E^{++}, T_2^{++})/a=\sqrt{3}$ and $r_1(A_1^{-+})/a=2$ on the 48I lattice. On the 64I lattice, these $r_1/a$'s are chosen to be $\sqrt{3}$, $\sqrt{5}$, $2\sqrt{2}$. At these values of $r_1$, $\Phi_2(r)$ are seen to be approximately zero. The effective mass of each $\mathcal{C}_{An}(r_1,t)$ for all the four channels on the two ensembles (48I and 64I) are illustrated in Fig.~\ref{fig:final_mass_plat}, where the vertical axis and the horizontal 
 axis are plotted in physical units. It is seen that the effective masses 
 show more or less plateaus at first several time slices, which indicate that $C_{A1}(r_1,t)$ can be described with a single exponential as expected. 
With the prescription mentioned above, the ground state masses in each channels are extracted through single exponential function forms in early time windows. In each panel of Fig.~\ref{fig:final_mass_plat}, the grey bands illustrate the fitted results and the time interval of the fits. In order to exhibit the efficacy of the CDER method, we also show the lattice data that are derived by calculating the correlation functions without the implementation of CDER in green (48I ensemble) and yellow bands (64I ensemble). Obviously, the data without CDER fluctuate drastically with respect to time and the errors are quite large. In contrast, the data points based on CDER method have much smaller errors and can be described by the single exponential function forms. It is impressive that the effective mass plateau of the $A^{++}_1$ ground state lasts over 5 time slices. 
The fitted mass values are listed in Table~\ref{tab:mass_final_results} in physical units.
We have the following observations 
\begin{itemize}
    \item In all the four channels involved, namely, $A_1^{++}, E^{++}, T_2^{++}$ and $A_1^{-+}$ channels, the ground state masses on $48I$ and $64I$ ensembles are consistent with each other within errors. Since the two ensembles have similar physical volumes but different lattice spacings, this consistence imply that the finite lattice spacing $a$ effects are not large.
    \item The masses of ground states in $E^{++}$ and $T_2^{++}$ channels are compatible with each other on the same ensemble. This also indicates that the discretization effects are not important since these states correspond to the $2^{++}$ state in the continuum limit.
    \item  If the ground states observed are free from excited states, their masses are slightly higher than but not far from those of the quenched approximation predictions and preliminary full-QCD calculations~\cite{Morningstar:1999rf,Chen:2005mg,Richards:2010ck,Gregory:2012hu,Sun:2017ipk}.
\end{itemize}

\begin{table}
	\caption{\label{tab:mass_final_results} Fitted masses of the ground states in $A_1^{++}, E^{++}, T_2^{++}, A_1^{-+}$ channels on 48I and 64I ensembles. The masses are converted into values in physical units (GeV) using the lattice spacings on Table~\ref{tab:lattice}.}
	\begin{ruledtabular}
		\begin{tabular}{ccccc}
			\texttt{ensemble} & $m(A_1^{++})$ & $m(E^{++})$ & $m(T_2^{++})$ & $m(A_1^{-+})$\\
			48I & $1.82 \pm 0.09$ & $2.6 \pm 0.2$ & $2.7 \pm 0.2$ & $2.8 \pm 0.2$ \\
			64I & $1.96 \pm 0.08$ & $2.7 \pm 0.1$ & $2.7 \pm 0.2$ & $2.8 \pm 0.2$\\
		\end{tabular}
	\end{ruledtabular}
\end{table} 
It should be noted that, the situation for glueballs is much complicated in the presence 
of dynamical quarks. Experimentally, most of hadrons are observed as resonances, therefore their decays should be taken into account. The energy levels obtained on the lattice are
actually the eigen energies of the lattice Hamiltonian, and their connection with the resonances in the real world is highly nontrivial. In full QCD, glueballs can mix with $\bar{q}q$ and multi-hadron states. However, with the gluonic operators, the $\bar{q}q$ states are suppressed by $\mathcal{O}(1/N_c)$ and two meson states are suppressed by $\mathcal{O}(1/N_c^2)$. This is perhaps the reason that we could see the effective mass palteaus in short time saparation. To have a complete picture, it is essential to include quark operators in the calculation, which is beyond the scope of the present work. There is an exploratory lattice study in this direction in the scalar channel~\cite{Brett:2019tzr}, but no glueball states can be identifiable unambiguously yet. 


\textit{Summary}
 An exploratory study of glueballs is performed on two $N_f=2+1$ gauge ensembles with large lattice sizes and the dynamical quark masses being tuned at the physical point. The large lattice size enables us to use the CDER technique to reduce the errors of the correlation functions of glueball operators. By using the spatially extended glueball operators defined through the gauge potential $A_\mu(x)$ in the Coulomb gauge, the Bethe-Salpeter wave functions of the scalar ($A_1^{++})$, the tensor ($E_2^{++}\oplus T_2^{++}$) and the pseudoscalar ($A_1^{-+})$ states are derived, which are similar to the features of the non-relativistic wave functions of two-body systems. Even though the physical meaning of these BS wave functions may not be clear, we can make use of their radial behaviors to further improve the ground state domination of the related correlation functions at the early time region, where the ground state masses in each channel can be extracted through single-exponential function form. To the extent that the excited state contamination is small and the couplings to $\bar{q}q$ meson states are negligible (they could show
 up at much large time separation given enough statistics). Our calculation with physical dynamical quarks turns out
 to be compatible with the quenched lattice results. A scrutinized full-QCD study should be carried out by including the $\bar{q}q$ and multi-meson operators, as well as considering the resonance nature of most hadrons. As a noise reduction scheme, the same CDER technique can be applied to all the correlation functions involving the disconnected quark insertions. In this sense, our study indicates a breakthrough to improve the signal-to-noise ratios of correlation functions involving glueball operators, and pave the way to the final answer on the existence of the glueball states.

\begin{acknowledgements}
We thank the RBC and UKQCD collaborations for providing us their DWF gauge configurations. This work is supported by the Strategic Priority Research Program of Chinese Academy of Sciences (No. XDB34030300 and XDPB15), the National Key Research and Development Program of China (No. 2020YFA0406400), and the National Natural Science Foundation of China (NNSFC) under Grants No.11935017, No.12070131001 (CRC 110 by DFG and NNSFC). The computations were performed on the CAS Xiandao-1 computing environment, the HPC clusters at Institute of High Energy Physics (Beijing), China Spallation Neutron Source (Dongguan) and Institute of Theoretical Physics. Y. Yang is also supported in part by a NSFC-DFG joint grant under grant No.12061131006 and SCHA~458/22.
\end{acknowledgements}

\bibliography{ref}

\begin{thebibliography}{16}%
\makeatletter
\providecommand \@ifxundefined [1]{%
 \@ifx{#1\undefined}
}%
\providecommand \@ifnum [1]{%
 \ifnum #1\expandafter \@firstoftwo
 \else \expandafter \@secondoftwo
 \fi
}%
\providecommand \@ifx [1]{%
 \ifx #1\expandafter \@firstoftwo
 \else \expandafter \@secondoftwo
 \fi
}%
\providecommand \natexlab [1]{#1}%
\providecommand \enquote  [1]{``#1''}%
\providecommand \bibnamefont  [1]{#1}%
\providecommand \bibfnamefont [1]{#1}%
\providecommand \citenamefont [1]{#1}%
\providecommand \href@noop [0]{\@secondoftwo}%
\providecommand \href [0]{\begingroup \@sanitize@url \@href}%
\providecommand \@href[1]{\@@startlink{#1}\@@href}%
\providecommand \@@href[1]{\endgroup#1\@@endlink}%
\providecommand \@sanitize@url [0]{\catcode `\\12\catcode `\$12\catcode
  `\&12\catcode `\#12\catcode `\^12\catcode `\_12\catcode `\%12\relax}%
\providecommand \@@startlink[1]{}%
\providecommand \@@endlink[0]{}%
\providecommand \url  [0]{\begingroup\@sanitize@url \@url }%
\providecommand \@url [1]{\endgroup\@href {#1}{\urlprefix }}%
\providecommand \urlprefix  [0]{URL }%
\providecommand \Eprint [0]{\href }%
\providecommand \doibase [0]{http://dx.doi.org/}%
\providecommand \selectlanguage [0]{\@gobble}%
\providecommand \bibinfo  [0]{\@secondoftwo}%
\providecommand \bibfield  [0]{\@secondoftwo}%
\providecommand \translation [1]{[#1]}%
\providecommand \BibitemOpen [0]{}%
\providecommand \bibitemStop [0]{}%
\providecommand \bibitemNoStop [0]{.\EOS\space}%
\providecommand \EOS [0]{\spacefactor3000\relax}%
\providecommand \BibitemShut  [1]{\csname bibitem#1\endcsname}%
\let\auto@bib@innerbib\@empty
\bibitem [{\citenamefont {Morningstar}\ and\ \citenamefont
  {Peardon}(1999)}]{Morningstar:1999rf}%
  \BibitemOpen
  \bibfield  {author} {\bibinfo {author} {\bibfnamefont {C.~J.}\ \bibnamefont
  {Morningstar}}\ and\ \bibinfo {author} {\bibfnamefont {M.~J.}\ \bibnamefont
  {Peardon}},\ }\href {\doibase 10.1103/PhysRevD.60.034509} {\bibfield
  {journal} {\bibinfo  {journal} {Phys. Rev. D}\ }\textbf {\bibinfo {volume}
  {60}},\ \bibinfo {pages} {034509} (\bibinfo {year} {1999})},\ \Eprint
  {http://arxiv.org/abs/hep-lat/9901004} {arXiv:hep-lat/9901004} \BibitemShut
  {NoStop}%
\bibitem [{\citenamefont {Chen}\ \emph {et~al.}(2006)\citenamefont {Chen} \emph
  {et~al.}}]{Chen:2005mg}%
  \BibitemOpen
  \bibfield  {author} {\bibinfo {author} {\bibfnamefont {Y.}~\bibnamefont
  {Chen}} \emph {et~al.},\ }\href {\doibase 10.1103/PhysRevD.73.014516}
  {\bibfield  {journal} {\bibinfo  {journal} {Phys. Rev. D}\ }\textbf {\bibinfo
  {volume} {73}},\ \bibinfo {pages} {014516} (\bibinfo {year} {2006})},\
  \Eprint {http://arxiv.org/abs/hep-lat/0510074} {arXiv:hep-lat/0510074}
  \BibitemShut {NoStop}%
\bibitem [{\citenamefont {Richards}\ \emph {et~al.}(2010)\citenamefont
  {Richards}, \citenamefont {Irving}, \citenamefont {Gregory},\ and\
  \citenamefont {McNeile}}]{Richards:2010ck}%
  \BibitemOpen
  \bibfield  {author} {\bibinfo {author} {\bibfnamefont {C.~M.}\ \bibnamefont
  {Richards}}, \bibinfo {author} {\bibfnamefont {A.~C.}\ \bibnamefont
  {Irving}}, \bibinfo {author} {\bibfnamefont {E.~B.}\ \bibnamefont {Gregory}},
  \ and\ \bibinfo {author} {\bibfnamefont {C.}~\bibnamefont {McNeile}}
  (\bibinfo {collaboration} {UKQCD Collaboration}),\ }\href {\doibase
  10.1103/PhysRevD.82.034501} {\bibfield  {journal} {\bibinfo  {journal} {Phys.
  Rev. D}\ }\textbf {\bibinfo {volume} {82}},\ \bibinfo {pages} {034501}
  (\bibinfo {year} {2010})},\ \Eprint {http://arxiv.org/abs/1005.2473}
  {arXiv:1005.2473 [hep-lat]} \BibitemShut {NoStop}%
\bibitem [{\citenamefont {Gregory}\ \emph {et~al.}(2012)\citenamefont
  {Gregory}, \citenamefont {Irving}, \citenamefont {Lucini}, \citenamefont
  {McNeile}, \citenamefont {Rago}, \citenamefont {Richards},\ and\
  \citenamefont {Rinaldi}}]{Gregory:2012hu}%
  \BibitemOpen
  \bibfield  {author} {\bibinfo {author} {\bibfnamefont {E.}~\bibnamefont
  {Gregory}}, \bibinfo {author} {\bibfnamefont {A.}~\bibnamefont {Irving}},
  \bibinfo {author} {\bibfnamefont {B.}~\bibnamefont {Lucini}}, \bibinfo
  {author} {\bibfnamefont {C.}~\bibnamefont {McNeile}}, \bibinfo {author}
  {\bibfnamefont {A.}~\bibnamefont {Rago}}, \bibinfo {author} {\bibfnamefont
  {C.}~\bibnamefont {Richards}}, \ and\ \bibinfo {author} {\bibfnamefont
  {E.}~\bibnamefont {Rinaldi}},\ }\href {\doibase 10.1007/JHEP10(2012)170}
  {\bibfield  {journal} {\bibinfo  {journal} {JHEP}\ }\textbf {\bibinfo
  {volume} {10}},\ \bibinfo {pages} {170} (\bibinfo {year} {2012})},\ \Eprint
  {http://arxiv.org/abs/1208.1858} {arXiv:1208.1858 [hep-lat]} \BibitemShut
  {NoStop}%
\bibitem [{\citenamefont {Sun}\ \emph {et~al.}(2018{\natexlab{a}})\citenamefont
  {Sun}, \citenamefont {Gui}, \citenamefont {Chen}, \citenamefont {Gong},
  \citenamefont {Liu}, \citenamefont {Liu}, \citenamefont {Liu}, \citenamefont
  {Ma},\ and\ \citenamefont {Zhang}}]{Sun:2017ipk}%
  \BibitemOpen
  \bibfield  {author} {\bibinfo {author} {\bibfnamefont {W.}~\bibnamefont
  {Sun}}, \bibinfo {author} {\bibfnamefont {L.-C.}\ \bibnamefont {Gui}},
  \bibinfo {author} {\bibfnamefont {Y.}~\bibnamefont {Chen}}, \bibinfo {author}
  {\bibfnamefont {M.}~\bibnamefont {Gong}}, \bibinfo {author} {\bibfnamefont
  {C.}~\bibnamefont {Liu}}, \bibinfo {author} {\bibfnamefont {Y.-B.}\
  \bibnamefont {Liu}}, \bibinfo {author} {\bibfnamefont {Z.}~\bibnamefont
  {Liu}}, \bibinfo {author} {\bibfnamefont {J.-P.}\ \bibnamefont {Ma}}, \ and\
  \bibinfo {author} {\bibfnamefont {J.-B.}\ \bibnamefont {Zhang}},\ }\href
  {\doibase 10.1088/1674-1137/42/9/093103} {\bibfield  {journal} {\bibinfo
  {journal} {Chin. Phys. C}\ }\textbf {\bibinfo {volume} {42}},\ \bibinfo
  {pages} {093103} (\bibinfo {year} {2018}{\natexlab{a}})},\ \Eprint
  {http://arxiv.org/abs/1702.08174} {arXiv:1702.08174 [hep-lat]} \BibitemShut
  {NoStop}%
\bibitem [{\citenamefont {Sun}\ \emph {et~al.}(2018{\natexlab{b}})\citenamefont
  {Sun}, \citenamefont {Alexandru}, \citenamefont {Chen}, \citenamefont
  {Draper}, \citenamefont {Liu},\ and\ \citenamefont {Yang}}]{Sun:2015enu}%
  \BibitemOpen
  \bibfield  {author} {\bibinfo {author} {\bibfnamefont {W.}~\bibnamefont
  {Sun}}, \bibinfo {author} {\bibfnamefont {A.}~\bibnamefont {Alexandru}},
  \bibinfo {author} {\bibfnamefont {Y.}~\bibnamefont {Chen}}, \bibinfo {author}
  {\bibfnamefont {T.}~\bibnamefont {Draper}}, \bibinfo {author} {\bibfnamefont
  {Z.}~\bibnamefont {Liu}}, \ and\ \bibinfo {author} {\bibfnamefont {Y.-B.}\
  \bibnamefont {Yang}} (\bibinfo {collaboration} {\ensuremath{\chi}QCD}),\
  }\href {\doibase 10.1088/1674-1137/42/6/063102} {\bibfield  {journal}
  {\bibinfo  {journal} {Chin. Phys. C}\ }\textbf {\bibinfo {volume} {42}},\
  \bibinfo {pages} {063102} (\bibinfo {year} {2018}{\natexlab{b}})},\ \Eprint
  {http://arxiv.org/abs/1507.02541} {arXiv:1507.02541 [hep-ph]} \BibitemShut
  {NoStop}%
\bibitem [{\citenamefont {Liu}\ \emph {et~al.}(2018)\citenamefont {Liu},
  \citenamefont {Liang},\ and\ \citenamefont {Yang}}]{Liu:2017man}%
  \BibitemOpen
  \bibfield  {author} {\bibinfo {author} {\bibfnamefont {K.-F.}\ \bibnamefont
  {Liu}}, \bibinfo {author} {\bibfnamefont {J.}~\bibnamefont {Liang}}, \ and\
  \bibinfo {author} {\bibfnamefont {Y.-B.}\ \bibnamefont {Yang}},\ }\href
  {\doibase 10.1103/PhysRevD.97.034507} {\bibfield  {journal} {\bibinfo
  {journal} {Phys. Rev. D}\ }\textbf {\bibinfo {volume} {97}},\ \bibinfo
  {pages} {034507} (\bibinfo {year} {2018})},\ \Eprint
  {http://arxiv.org/abs/1705.06358} {arXiv:1705.06358 [hep-lat]} \BibitemShut
  {NoStop}%
\bibitem [{\citenamefont {Yang}\ \emph {et~al.}(2018)\citenamefont {Yang},
  \citenamefont {Gong}, \citenamefont {Liang}, \citenamefont {Lin},
  \citenamefont {Liu}, \citenamefont {Pefkou},\ and\ \citenamefont
  {Shanahan}}]{Yang:2018bft}%
  \BibitemOpen
  \bibfield  {author} {\bibinfo {author} {\bibfnamefont {Y.-B.}\ \bibnamefont
  {Yang}}, \bibinfo {author} {\bibfnamefont {M.}~\bibnamefont {Gong}}, \bibinfo
  {author} {\bibfnamefont {J.}~\bibnamefont {Liang}}, \bibinfo {author}
  {\bibfnamefont {H.-W.}\ \bibnamefont {Lin}}, \bibinfo {author} {\bibfnamefont
  {K.-F.}\ \bibnamefont {Liu}}, \bibinfo {author} {\bibfnamefont
  {D.}~\bibnamefont {Pefkou}}, \ and\ \bibinfo {author} {\bibfnamefont
  {P.}~\bibnamefont {Shanahan}},\ }\href {\doibase 10.1103/PhysRevD.98.074506}
  {\bibfield  {journal} {\bibinfo  {journal} {Phys. Rev. D}\ }\textbf {\bibinfo
  {volume} {98}},\ \bibinfo {pages} {074506} (\bibinfo {year} {2018})},\
  \Eprint {http://arxiv.org/abs/1805.00531} {arXiv:1805.00531 [hep-lat]}
  \BibitemShut {NoStop}%
\bibitem [{\citenamefont {Blum}\ \emph {et~al.}(2016)\citenamefont {Blum} \emph
  {et~al.}}]{RBC:2014ntl}%
  \BibitemOpen
  \bibfield  {author} {\bibinfo {author} {\bibfnamefont {T.}~\bibnamefont
  {Blum}} \emph {et~al.} (\bibinfo {collaboration} {RBC, UKQCD}),\ }\href
  {\doibase 10.1103/PhysRevD.93.074505} {\bibfield  {journal} {\bibinfo
  {journal} {Phys. Rev. D}\ }\textbf {\bibinfo {volume} {93}},\ \bibinfo
  {pages} {074505} (\bibinfo {year} {2016})},\ \Eprint
  {http://arxiv.org/abs/1411.7017} {arXiv:1411.7017 [hep-lat]} \BibitemShut
  {NoStop}%
\bibitem [{sm()}]{sm}%
  \BibitemOpen
  \href@noop {} {\bibinfo  {journal} {Supplementary materials}\ }\BibitemShut
  {NoStop}%
\bibitem [{\citenamefont {de~Forcrand}\ and\ \citenamefont
  {Liu}(1992)}]{deForcrand:1991kc}%
  \BibitemOpen
\bibfield  {journal} {  }\bibfield  {author} {\bibinfo {author} {\bibfnamefont
  {P.}~\bibnamefont {de~Forcrand}}\ and\ \bibinfo {author} {\bibfnamefont
  {K.-F.}\ \bibnamefont {Liu}},\ }\href {\doibase 10.1103/PhysRevLett.69.245}
  {\bibfield  {journal} {\bibinfo  {journal} {Phys. Rev. Lett.}\ }\textbf
  {\bibinfo {volume} {69}},\ \bibinfo {pages} {245} (\bibinfo {year}
  {1992})}\BibitemShut {NoStop}%
\bibitem [{\citenamefont {Liang}\ \emph {et~al.}(2015)\citenamefont {Liang},
  \citenamefont {Chen}, \citenamefont {Chiu}, \citenamefont {Gui},
  \citenamefont {Gong},\ and\ \citenamefont {Liu}}]{Liang:2014jta}%
  \BibitemOpen
  \bibfield  {author} {\bibinfo {author} {\bibfnamefont {J.}~\bibnamefont
  {Liang}}, \bibinfo {author} {\bibfnamefont {Y.}~\bibnamefont {Chen}},
  \bibinfo {author} {\bibfnamefont {W.-F.}\ \bibnamefont {Chiu}}, \bibinfo
  {author} {\bibfnamefont {L.-C.}\ \bibnamefont {Gui}}, \bibinfo {author}
  {\bibfnamefont {M.}~\bibnamefont {Gong}}, \ and\ \bibinfo {author}
  {\bibfnamefont {Z.}~\bibnamefont {Liu}},\ }\href {\doibase
  10.1103/PhysRevD.91.054513} {\bibfield  {journal} {\bibinfo  {journal} {Phys.
  Rev. D}\ }\textbf {\bibinfo {volume} {91}},\ \bibinfo {pages} {054513}
  (\bibinfo {year} {2015})},\ \Eprint {http://arxiv.org/abs/1411.1083}
  {arXiv:1411.1083 [hep-lat]} \BibitemShut {NoStop}%
\bibitem [{\citenamefont {Salpeter}\ and\ \citenamefont
  {Bethe}(1951)}]{bib:bethe_salpeter}%
  \BibitemOpen
  \bibfield  {author} {\bibinfo {author} {\bibfnamefont {E.~E.}\ \bibnamefont
  {Salpeter}}\ and\ \bibinfo {author} {\bibfnamefont {H.~A.}\ \bibnamefont
  {Bethe}},\ }\href {\doibase 10.1103/PhysRev.84.1232} {\bibfield  {journal}
  {\bibinfo  {journal} {Phys. Rev.}\ }\textbf {\bibinfo {volume} {84}},\
  \bibinfo {pages} {1232} (\bibinfo {year} {1951})}\BibitemShut {NoStop}%
\bibitem [{\citenamefont {Souza}\ \emph {et~al.}(2020)\citenamefont {Souza},
  \citenamefont {Narciso~Ferreira}, \citenamefont {Aguilar}, \citenamefont
  {Papavassiliou}, \citenamefont {Roberts},\ and\ \citenamefont
  {Xu}}]{Souza:2019ylx}%
  \BibitemOpen
  \bibfield  {author} {\bibinfo {author} {\bibfnamefont {E.~V.}\ \bibnamefont
  {Souza}}, \bibinfo {author} {\bibfnamefont {M.}~\bibnamefont
  {Narciso~Ferreira}}, \bibinfo {author} {\bibfnamefont {A.~C.}\ \bibnamefont
  {Aguilar}}, \bibinfo {author} {\bibfnamefont {J.}~\bibnamefont
  {Papavassiliou}}, \bibinfo {author} {\bibfnamefont {C.~D.}\ \bibnamefont
  {Roberts}}, \ and\ \bibinfo {author} {\bibfnamefont {S.-S.}\ \bibnamefont
  {Xu}},\ }\href {\doibase 10.1140/epja/s10050-020-00041-y} {\bibfield
  {journal} {\bibinfo  {journal} {Eur. Phys. J. A}\ }\textbf {\bibinfo {volume}
  {56}},\ \bibinfo {pages} {25} (\bibinfo {year} {2020})},\ \Eprint
  {http://arxiv.org/abs/1909.05875} {arXiv:1909.05875 [nucl-th]} \BibitemShut
  {NoStop}%
\bibitem [{\citenamefont {Brett}\ \emph {et~al.}(2020)\citenamefont {Brett},
  \citenamefont {Bulava}, \citenamefont {Darvish}, \citenamefont {Fallica},
  \citenamefont {Hanlon}, \citenamefont {H\"orz},\ and\ \citenamefont
  {Morningstar}}]{Brett:2019tzr}%
  \BibitemOpen
  \bibfield  {author} {\bibinfo {author} {\bibfnamefont {R.}~\bibnamefont
  {Brett}}, \bibinfo {author} {\bibfnamefont {J.}~\bibnamefont {Bulava}},
  \bibinfo {author} {\bibfnamefont {D.}~\bibnamefont {Darvish}}, \bibinfo
  {author} {\bibfnamefont {J.}~\bibnamefont {Fallica}}, \bibinfo {author}
  {\bibfnamefont {A.}~\bibnamefont {Hanlon}}, \bibinfo {author} {\bibfnamefont
  {B.}~\bibnamefont {H\"orz}}, \ and\ \bibinfo {author} {\bibfnamefont
  {C.}~\bibnamefont {Morningstar}},\ }\href {\doibase 10.1063/5.0008566}
  {\bibfield  {journal} {\bibinfo  {journal} {AIP Conf. Proc.}\ }\textbf
  {\bibinfo {volume} {2249}},\ \bibinfo {pages} {030032} (\bibinfo {year}
  {2020})},\ \Eprint {http://arxiv.org/abs/1909.07306} {arXiv:1909.07306
  [hep-lat]} \BibitemShut {NoStop}%
\bibitem [{\citenamefont {Liang}\ \emph {et~al.}(2014)\citenamefont {Liang},
  \citenamefont {Chen}, \citenamefont {Gong}, \citenamefont {Gui},
  \citenamefont {Liu}, \citenamefont {Liu},\ and\ \citenamefont
  {Yang}}]{Liang:2013eoa}%
  \BibitemOpen
  \bibfield  {author} {\bibinfo {author} {\bibfnamefont {J.}~\bibnamefont
  {Liang}}, \bibinfo {author} {\bibfnamefont {Y.}~\bibnamefont {Chen}},
  \bibinfo {author} {\bibfnamefont {M.}~\bibnamefont {Gong}}, \bibinfo {author}
  {\bibfnamefont {L.-C.}\ \bibnamefont {Gui}}, \bibinfo {author} {\bibfnamefont
  {K.-F.}\ \bibnamefont {Liu}}, \bibinfo {author} {\bibfnamefont
  {Z.}~\bibnamefont {Liu}}, \ and\ \bibinfo {author} {\bibfnamefont {Y.-B.}\
  \bibnamefont {Yang}},\ }\href {\doibase 10.1103/PhysRevD.89.094507}
  {\bibfield  {journal} {\bibinfo  {journal} {Phys. Rev. D}\ }\textbf {\bibinfo
  {volume} {89}},\ \bibinfo {pages} {094507} (\bibinfo {year} {2014})},\
  \Eprint {http://arxiv.org/abs/1310.3532} {arXiv:1310.3532 [hep-lat]}
  \BibitemShut {NoStop}%
\end{thebibliography}%
\begin{widetext}

\section*{Supplementary materials}
\subsection*{S1. Effective mass of the correlation function $\mathcal{C}_1(t)$}
By solving the generalized eigenvalue problem $\mathcal{C}_{\alpha\beta}(t_0)v_\beta=\lambda \mathcal{C}_{\alpha\beta}(0)v_\beta$, we can obtain the optimized operator $\mathcal{O}_n=v^{(n)}_\alpha \mathcal{O}_\alpha$ that is expected to couple most to the $n$-th state in a specifc $R^{PC}$ channel, where $v_\alpha^{(n)}$ is the eigenvector corresponding to the $n$-th largest eigenvalue $\lambda^{(n)}$. Thus the correlation function 
\begin{equation}
    \mathcal{C}_{1}(t)=\frac{1}{T}\sum\limits_\tau \langle 0|\mathcal{O}_n(t+\tau)\mathcal{O}_n(\tau)|0\rangle
\end{equation}
should be dominated by the contribution from the ground state. After introducing the 
effective mass function 
\begin{equation}
    m_\mathrm{eff}(t)=\log \frac{\mathcal{C}_1(t)}{\mathcal{C}_1(t+1)},
\end{equation}
the ground state dominance can be monitored by the temporal behavior of $ m_\mathrm{eff}(t)$. Figure~\ref{fig:GG_mass} shows $m_\mathrm{eff}(t)$ of $A_1^{++}$,
$A_1^{-+}$, $E^{++}$ and $T_2^{++}$ channels on ensemble 48I. It is seen that for each of these channels, $m_\mathrm{eff}(t)$ has good signals but appreciable time dependence at the first several time slices. This indicates that the contamination from higher states has not been suppressed drastically even with the optimized operator $\mathcal{O}_1$. 
\begin{figure}[t]
	\centering
	\includegraphics[height=10.0cm]{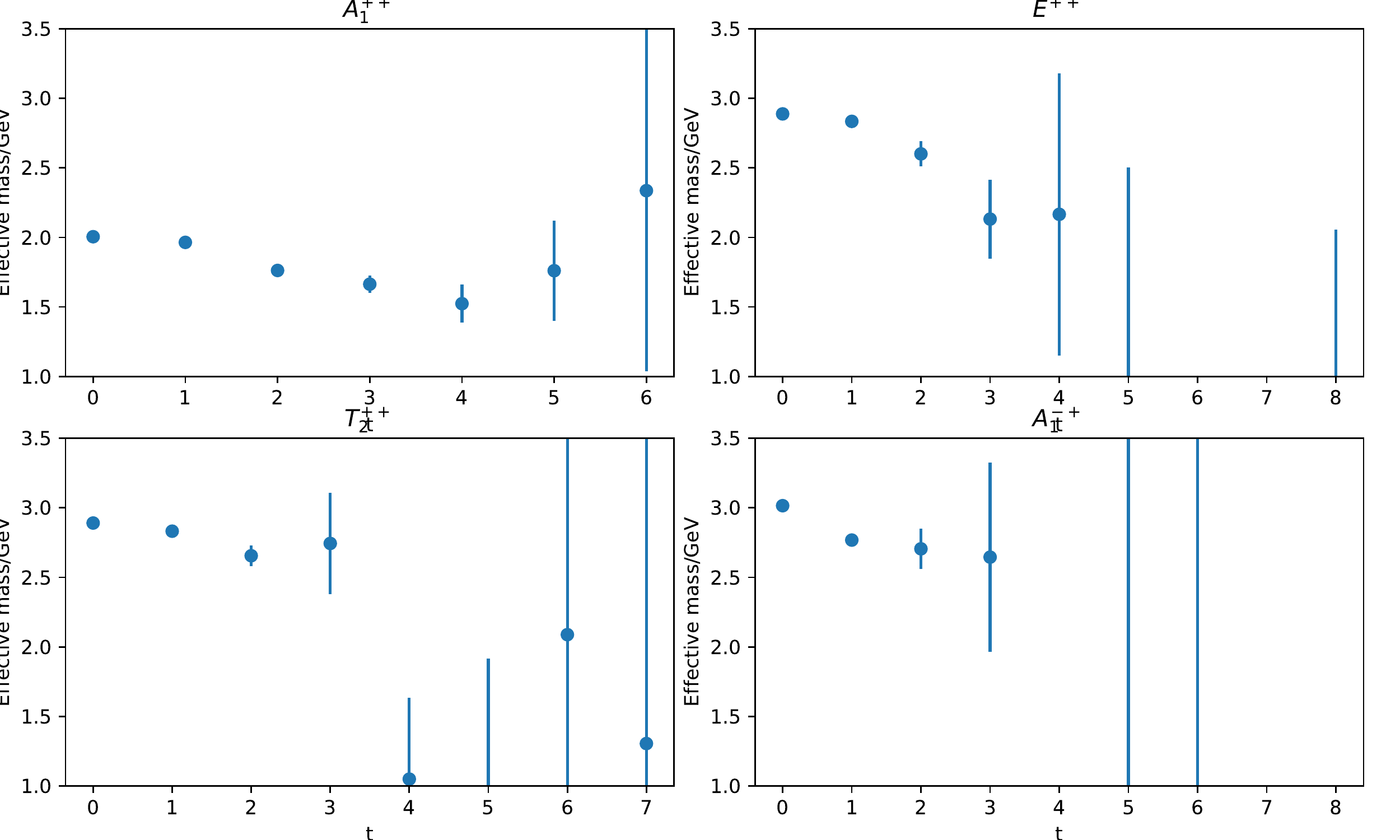}
	\caption{Effective mass plots of $\mathcal{C}_1(t)$ for $A_1^{++}$,
$A_1^{-+}$, $E^{++}$ and $T_2^{++}$ channels on 48I.}
	\label{fig:GG_mass}
\end{figure}

\subsection*{S2. Parameterization of BS functions}
 The BS wave functions in Fig.~\ref{fig:bs-wavefunc} exhibit similar features to the radial wave functions of a two-body system with a central potential, so we try to parameterize them tentatively using phenomenological functions forms. For the scalar ($A_1^{++}$) and the 
 tensor ($E^{++}$ and $T_2^{++}$) channels, the BS wave functions of the ground states and the first excited states, namely $\Phi_1(r)$ and $\Phi_2(r)$, are very similar to the $1S$ and $2S$ wave functions, therefore we describe them using the following function forms~\cite{Liang:2014jta}
\begin{eqnarray}\label{eq:s-wave}
\Phi_1(r)&=&\Phi_1(0) e^{-(r/r_0)^\alpha}+C\nonumber\\
\Phi_2(r)&=&\Phi_2(0) (1+\beta r^\alpha) e^{-(r/r_0)^\alpha}+C'
\end{eqnarray}
where the constant terms $C$ and $C'$ account for the constant tails in the large $r$ region, which may be due to the unphysical modes of the domain-wall sea quarks~\cite{Liang:2013eoa}. For the pseudoscalar $A_1^{-+}$ channel, $\Phi_{1,2}(r)$ show the typical feature $\Phi_{1,2}(0)=0$ of $P$-wave radial wave functions. Therefore, the functions can be parameterized as
\begin{eqnarray}\label{eq:p-wave}
\Phi_1(r) & =& A~r~e^{-(r / r_0)^\alpha}+D\nonumber\\
\Phi_2(r) & =& A~r~(1 - \beta r^\alpha) e^{-(r / r_0)^\alpha}+D'
\end{eqnarray}

We perform naive fits to $\Phi_{1,2}(r)$ in each channel 
using the corresponding function forms mentioned above
The fitted parameter values are shown in Table~\ref{tab:bs_func_fitting_params}. It should be noted that the lineshapes and the parameters of BS wave function in the scalar and tensor channels (except for the constant terms) are approximately agree with those 
from those in quenched approximation~\cite{Liang:2014jta}. The $\Phi_{1,2}(r)$ for the pseudoscalar has been obtained for the first times on the lattice. These BS wave functions may imply that the scalar and tensor glueballs can be viewed as 
two-constituent-gluon systems in $S$-wave, while the pseudoscalar glueballs are two-constituent-gluon systems in $P$-wave. Of course these are phenomenological interpretations in the picture of the constituent model. 
\begin{table}
	\caption{\label{tab:bs_func_fitting_params} Fitted parameters of Bethe-Salpeter wave function form in Eq.~(\ref{eq:s-wave}) and (\ref{eq:p-wave}).}
	\begin{ruledtabular}
		\begin{tabular}{cccc}
			    & $r_0$ & $\alpha$ & $\beta$ \\
			48I $A_1^{++}$ level 1 & $0.14 \pm 0.04$ & $1.6 \pm 0.2$ & \\
			48I $A_1^{++}$ level 2 & $0.16 \pm 0.05$ & $1.5 \pm 0.2$ & $24 \pm 1$\\
			64I $A_1^{++}$ level 1 & $0.15 \pm 0.03$ & $1.4 \pm 0.1$ & \\
			64I $A_1^{++}$ level 2 & $0.20 \pm 0.10$ & $1.5 \pm 0.4$ & $18 \pm 2$\\
			48I $E^{++}$ level 1   & $0.24 \pm 0.07$ & $1.8 \pm 0.2$ & \\
			48I $E^{++}$ level 2   & $0.22 \pm 0.06$ & $1.6 \pm 0.2$ & $16 \pm 1$\\
			64I $E^{++}$ level 1   & $0.24 \pm 0.06$ & $1.7 \pm 0.2$ & \\
			64I $E^{++}$ level 2   & $0.22 \pm 0.08$ & $1.7 \pm 0.3$ & $17 \pm 2$\\
			48I $T_2^{++}$ level 1 & $0.26 \pm 0.09$ & $2.0 \pm 0.3$ & \\
			48I $T_2^{++}$ level 2 & $0.25 \pm 0.08$ & $1.8 \pm 0.3$ & $19 \pm 2$\\
			64I $T_2^{++}$ level 1 & $0.26 \pm 0.06$ & $1.9 \pm 0.2$ & \\
			64I $T_2^{++}$ level 2 & $0.23 \pm 0.05$ & $1.8 \pm 0.2$ & $18 \pm 1$\\
			48I $A_1^{-+}$ level 1 & $0.24 \pm 0.08$ & $1.9 \pm 0.3$ &\\
			48I $A_1^{-+}$ level 2 & $0.23 \pm 0.07$ & $1.8 \pm 0.3$ & $12 \pm 1$\\
			64I $A_1^{-+}$ level 1 & $0.20 \pm 0.05$ & $1.7 \pm 0.2$ &\\
			64I $A_1^{-+}$ level 2 & $0.19 \pm 0.06$ & $1.6 \pm 0.2$ & $10.1 \pm 0.7$\\
		\end{tabular}
	\end{ruledtabular}
\end{table}

\end{widetext}

\end{document}